\documentstyle[12pt,fleqn]{article}
\newcommand{\rf}[1]{\raisebox{1ex}{\cite{#1}}}

\newcommand{\fg}[1]{\item {\label{#1}}}

\newcommand{\barr}{\begin{array}}
\newcommand{\ear}{\end{array}}
\newcommand{\bea}{\begin{eqnarray}}
\newcommand{\eea}{\end{eqnarray}}
\newcommand{\beq}{\begin{equation}}
\newcommand{\eeq}{\end{equation}}

\begin{document}


\vskip 40pt

\centerline {\bf Symbolic dynamics  I}

\centerline {\bf Finite dispersive billiards.}
\vskip 1.5cm                                                               
  
\centerline { 
Kai T. Hansen
           }
\centerline {\em Niels Bohr Institute
\footnote{
{\ddag}
{\small Permanent address: Phys Dep., University of Oslo, Box 1048, Blindern,
   N-0316 Oslo}
} 
}
\centerline {\em Blegdamsvej 17, DK-2100 Copenhagen \O}  
\centerline {{\em e-mail:}  khansen@nbivax.nbi.dk}

\vskip 25pt
{\centerline{\bf ABSTRACT}}
\noindent{
Orbits in different dispersive billiard systems, e.g. the 3  disk system,
are mapped into a topological well ordered symbol  plane and it is
showed that forbidden and allowed orbits are separated by  a monotone
pruning front.  The pruning front can be approximated by a sequence of
finite symbolic dynamics grammars.
   }
 
\newpage

\section{Introduction}

The aim of this article is to find an effective description of all 
allowed orbits in a classical billiard system.  The systems
considered are a point particle on a plane, 
moving freely and reflecting elastically 
off dispersing walls.  In most of our examples the walls are the borders to 
circular shaped reflecting disks.
Such systems have recently been in focus of much theoretical interest,
because of their utility in studies of the
quantum mechanics of classical chaotic systems \rf{cqc,gaspard,ss}.  
Semiclassical calculations of the quantal spectrum of 
classically chaotic systems require  summations 
over all periodic orbits of the classical
system\rf{gutzwiller}, and it is essential to have an effective
description of the allowed classical orbits.


The solution we propose here is to apply techniques developed in the study of
one dimensional unimodal maps, and of two dimensional
H\'enon-type maps.  
  In the unimodal maps the ordering of allowed orbits
along  the parameter axis is given by the MSS theory \rf{mss},
i.e. the kneading sequence of the
critical point\rf{mt}.  From the kneading sequence one constructs a binary
number $\gamma_c$
and  the interval $(\gamma_c,1)$ is 
the primary pruned region of the unimodal map.  An orbit is forbidden if
the orbit, or any of its images or preimages yields 
a binary number $\gamma$ in
the pruned region $(\gamma_c,1)$.    A convenient 
visualization of the allowed orbits is afforded by
a binary tree on which every allowed orbit
corresponds to a path down the tree.  The forbidden orbits correspond to
the forbidden branches, i.e. the branches that have
to be {\it pruned} \rf{cgp} from the tree of allowed orbits.

A unimodal map such as the logistic map has a parameter space window for each
stable periodic orbit.  Within such window
the kneading sequence is periodic,  one can find a finite
number of forbidden symbol sub-sequences, and  construct
a new unpruned alphabet\rf{cvitanovic}.   One can also
describe allowed sequences as walks on a finite Markov graph constructed from
the pruned binary tree\rf{grassberger}.  Chaotic dynamics
does not in general have a finite Markov partition,
but we believe\rf{predrag} that the statistical properties 
can be approximated from calculations on
repellers corresponding to converging sequences of nearby
windows with stable orbits. 
  
In ref.\rf{cgp} Cvitanovi\'c, 
Gunaratne and Procaccia have extended these methods to two 
dimensional maps like the H\'enon map\rf{henon} and showed that 
numerically one can obtain a connected primary
pruned region in a two dimensional space $(\gamma , \delta)$ constructed
from the properly ordered symbolic future and the past
of the orbits;  the border of the
area is the pruning front.  
The conjecture is that this curve is  monotone, and that  there are no 
other forbidden orbits
than those that pass through the primary pruning region.  
One can approximate the grammar of forbidden orbits by forbidding finite 
sequences that correspond to rectangles in
the symbol plane. Each such approximation gives a finite number of finite
length forbidden symbol 
sequences.  From these sequences one can make successively refined 
Markov partitions and obtain well converging estimates of averages 
such as the topological entropy\rf{politi,politi_grass,losal}.
The pruning front is the symbol plane representation of the
primary homoclinic tangencies introduced by Grassberger and Kantz\rf{gk} in 
order to  partition  the H\'enon map non-wandering set into a
``left'' and a ``right'' side.

\subsection{Symbols in billiard systems}

In billiard systems the choice of symbols is much easier; 
a symbol for bounce off
each concave wall, e.g. a disk or a branch of the hyperbola, 
is natural and unambiguous\rf{eck_win,freddy,cvi_eck,cvi_eck_sym}. 
Billiards considered here are hyperbolic systems with only unstable orbits; 
the stable and unstable manifolds of the orbits are never tangential to
each other.  An orbit in the billiard is unstable also at the point
where it is
pruned, in contrast to the H\'enon map and
smooth  Hamiltonian potentials, where pruning  always involves inverse
bifurcations and stable orbits \rf{DR_prl,DR}.

Construction of billiard symbolic dynamics proceeds in 2 steps:
\ 1. define a symbol plane that is topologically {\em well
ordered}; \ \ 2.  determine the {\em boundary
orbits} between the legal and the illegal orbits.  The well
ordered  symbol plane is defined by introducing a new symbol alphabet
and associating with every orbit a pair of coordinates constructed from 
the alphabet.  A similar well ordered  symbolic alphabet has been introduced
for a different scattering problem by Troll\rf{troll}.
The second issue is simple. 
We show that for the billiards one starts a tangential bounce from
every point on the wall and these are the boundary orbits.  

The symbol plane we construct is complete for a system with a finite number
of hard disks situated far from each other in the configuration space.
A physically motivated example is a particle bouncing between 4 
hyperbolas\rf{ss,hansen}.  This billiard do not have a obvious limit of a
complete Cantor set but we use the same symbol plane as the 4-disks 
billiard and orbits are pruned the same way.


\section{Covering alphabet for dispersing billiards}

For simplicity we choose to describe the covering alphabets for
dispersing billiards by using billiards with a finite number of
identical circularly shaped walls. In general the walls do not have to be
circular, but they do have to be dispersing.  When they are
identical, the configuration space symmetry enables us to
reduce the number of symbols and to simplify the problem.

If we take a billiard system with {\em sufficiently separated} disks 
in the plane there exist an orbit which visits the disks in any desired order.

Let the disks be enumerated by $s \in \{1,2,\ldots ,N\}$ and denote the
bounce of the point particle on disk $s$ at the discrete time $t$ by symbol
$s_t$.   Dispersing (concave) walls can not have two succeeding bounces
off the same 
wall, i.e.
the subsequence $\_ ss\_$ is forbidden, but all other sequences are allowed
for sufficiently separated disks.

Given the infinite symbolic past of the particle bouncing on disk
$s_0$ at time $t=0$, that is $\ldots s_{-3}s_{-2}s_{-1}$, and 
the infinite symbolic future $s_1s_2s_3\ldots$, the
position of the bounce $x$ and the outgoing angle $\phi$ measured from
the normal vector of the wall
are uniquely determined. The phase space coordinates we use are the
position and the angle $(x,\phi)$ of the bounce off disk $s_0$.

If the N-disk system is open, then almost all points $(x,\phi)$ 
correspond to a particle that has entered the system from outside at some
earlier time and is going to escape from the system at some later time.
The union of all points contained within the billiard for all
times is the non-wandering set $\Lambda$.  For a sufficiently 
separated disk
billiard, $\Lambda$ is a Cantor set in the phase space similar to the
non-wandering set for the
Smale horseshoe \rf{smale}.    Each point in this Cantor set is described
by a unique symbolic past and future.  However, if we use the disk
symbols $s_t$ to describe a point in the Cantor set, there is
no obvious ordering of the symbols from ``small'' symbols to 
``large'' symbols that would reflect an increase of the position
or angle coordinate in
the phase space.  A symbolic description with a natural ordering of this
kind we call well ordered symbols and we show how these can be
constructed by a few 
examples.  Other configurations of disks can be worked out in a
similar way.


\subsection{3 disks symbolic dynamics}

Let the billiard system be 3 equal disks with radius 1 
and the distance $r$ between the centers, enumerated anticlockwise
as in figure~\ref{f_3_config}.  
Without loss of generality we
assume that the particle bounces off disk number~1 at time $t=0$ ($s_0=1$). 
The position $x$ is the angle describing the bounce on disc 1 and 
$\phi $ is the angle from the normal vector.
Define $\Lambda^+_t$ as the union of points in phase space that
at least give $t$ bounces in the billiard after bouncing off disc~1. 
Thus $\Lambda^+_1$ is all $(x,\phi )$ of disk 1 corresponding to a particle 
that bounces either off disk 2 or off disk 3 at time $t=1$.
$\Lambda^+_1$ consists of 
two diagonal strips in phase space, with $\phi$ decreasing when 
$x$ is increasing. 
The lower strip in the phase space
is the union of all points in phase space corresponding to a particle 
bouncing off disk 2 at time $t=1$ and the 
symbol sequence for this strip is $s_0s_1 = 12$. The upper strip
is the union of all points corresponding to a particle that bounces 
off disk 3 at time $t=1$, 
and this strip has symbol sequence $s_0s_1=13$. 
The order of  the two strips with increasing $x$ or $\phi$ is 
$s_0s_1=\{12,13\}$. 
$\Lambda^+_2$ is four diagonal strips, two inside each of the
two strips of $\Lambda^+_1$.  The symbol description of the 
four strips ordered after increasing values of
$x$ or $\phi$ is $s_0s_1s_2 = \{ 121,123,132,131\} $.
Table~\ref{t_3_symbols} shows
the symbol sequences for 4 bounces ordered after increasing values of
$x$ and $\phi$. 
The limit of $\Lambda^+_t$ when $t\rightarrow \infty$ is a Cantor set
of lines with a unique labeling $s_0s_1s_2\ldots$ for each line.

Let new symbols be defined from a combination of two symbols $s_{t-1}$
and $s_t$, and such that the ordering of the Cantor set lines in the phase 
space is preserved.
In table~\ref{t_3_symbols} new symbols $w_t \in \{ 0,1\} $ are written
together with the old symbols for the first 4 bounces.  
>From the symbols $w_1w_2w_3\ldots$ a rational binary
number $\gamma = 0.w_1w_2w_3\ldots = \sum_{t=1}^{\infty} w_t / 2^t$
is constructed.  This symbolic coordinate $\gamma$ 
is increasing with increasing values of $x$ and increasing
values of $\phi$.  To construct the symbols $w_t$ let first $v_t$
denote if two  
consecutive bounces $s_{t-1}$ and $s_t$ take place in a clockwise or 
anticlockwise direction. The symbols $v_t$ are not well ordered because 
a bounce in a dispersing (concave) wall reverses the ordering in the phase
space.  At each bounce one has to  
invert the symbols $v_t$ and this gives the algorithm:
\begin{equation}  
 \begin{array}{lcr}
  v_t & = s_t-s_{t-1} &  \\
  \mbox{if} & v_t<1
     & \mbox{then}\;\;\; v_t = v_t + 3  \nonumber \\
  w_t & = &  \left\{ \begin{array}{l} 
			v_t - 1\;\;\; \mbox{if}\;\;\; t\mbox{\ odd} \\ 
			2 - v_t \;\;\; \mbox{if}\;\;\; t\mbox{\ even} \\ 
                   \end{array}  \right. \nonumber
 \end{array} \label{a_3_disk}
\end{equation}

The new symbols are constructed to reflect the ordering of the lines
$\Lambda^+_\infty$ in the phase space. 
Exchanging
{\em odd} and {\em even} in algorithm~(\ref{a_3_disk}) gives symbols
$w_t'=1-w_t$, and a binary number $\gamma'
= 0.w_1'w_2'w_3'\ldots$ decreasing with $x$ and $\theta$ and 
therefore also the symbols $w_t'$ are well ordered. 
>From a orbit described by symbols $s_t$ we have two mappings to the
two well ordered symbols $w_t$ and $w_t'$.

The particle bouncing at $(x,\phi)$ also has a past, and a symbol
sequence describing this past.  Let $\Lambda^-_t $ be 
the union of points in phase space corresponding to a particle that
has at least $t$ bounces in the billiard before arriving at
$(x,\phi)$.  Then $\Lambda^-_1$ is the union of all points arriving 
at disk 1 after being bounced off
disk 2 or off disk 3, and $\Lambda^-_1$ is two diagonal strips in the 
phase space.  The
incoming angle has opposite sign to the outgoing angle and the
strips in $\Lambda^-_1$ have increasing $\phi$ with increasing $x$.
We get the new symbols $w_t$ from algorithm (\ref{a_3_disk}) with $t\leq 0$.
The symbolic coordinate for the past is
$\delta = 0.w_0w_{-1}w_{-2}\ldots = \sum_{t=1}^{\infty} w_{1-t} /
2^t$.
The coordinate $\delta$ is increasing with increasing value of $x$
but $\delta$ is decreasing with increasing value of $\phi$.
If we choose the value $\gamma'$ for the future we have
$\delta' = 0.w_0'w_{-1}'w_{-2}'\ldots$ for the past.

Let $\Lambda_t = \Lambda^+_t \cap \Lambda^-_t$.
Then $\Lambda_2$ consist of
$16 = (2\cdot 2)^2$ areas in the phase space.
Each of the 16 disjunct sets of $\Lambda_2$
has a unique enumeration by the 4 binary symbols $\{ w_{-1},
w_{0}, w_{1},w_{2} \}$.  The set $\Lambda =\Lambda_{\infty}$ is a Cantor
set and each point in the set is represented by a bi-infinite symbol
string $\ldots w_{-2}w_{-1}w_{0}w_{1}w_{2}w_{3}\ldots $.
The set $\Lambda$ is the non--wandering set of the billiard.

The symbolic plane is the unit square  $(\gamma ,\delta
)$;  $0\leq \gamma ,\delta \leq 1$.  Each non--wandering orbit is
represented by a
point in this plane, and each point in the plane is one of the 
non--wandering orbits.  
As mentioned earlier, we actually get two points in the symbol
plane from one orbit.  Because of the symmetries of rotation and
reflection in the billiard, we get a number of different orbits in the
billiard from one point in the symbol plane, but all the orbits have 
the same length and stability.
The symbolic plane is a representation of the
phase space of the  well separated billiard, with all gaps in the
Cantor set removed.   This is a very convenient space  to work in;
the curved lines in the phase space become straight lines and
this space does not change if the distance $r$ changes or if
the borders of the disks are slightly changed as long as the
symmetry and concavity is kept.  

The billiard can be reduced to a fundamental domain\rf{cvi_eck_sym}.  
The fundamental domain is one sixth of the original 3 disk
billiard and it is tiling the whole billiard.  In our phase space the 
fundamental domain is the part with $x\leq \pi /6$ and is constructed
in the symbol plane as follows:  The symbol sequence $\ldots
s_{-2}s_{-1}s_{0}s_{1}s_{2}\ldots$  gives two symbolic coordinates
$(\gamma ,\delta )$ and $(\gamma' ,\delta' )$ but if $\gamma + \delta \neq
1$ then one point is in the fundamental region and the other
point is in the region $x > \pi /6$.   
We find the two points $(\gamma ,\delta )$ and $(\gamma' ,\delta' )$
and choose $(\gamma ,\delta )$  if $\gamma +\delta <1$ or choose
$(\gamma' ,\delta' )$ if $\gamma+\delta>1$.  If $\gamma + \delta =1$ then
both $(\gamma ,\delta )$ and $(\gamma' ,\delta' )$ are on the border
of the fundamental domain, and as our convention we choose $(\gamma
,\delta )$  if $\gamma > \frac{1}{2}$ and choose 
$(\gamma' ,\delta' )$ if $\gamma <  \frac{1}{2} $. 
If we have a billiard without symmetry we only use one map from $s_t$ to
$w_t$ and we do not have any fundamental domain.


\subsection{Symbolic dynamics for $N$ disks on a circle}

As a generalization of the 3 disk billiard let $N$ equal disks
have the center of each disk on a large circle and let the 
distance between centers of neighbor disks on the large circle
be $r$.  Then $\Lambda^+_1$
is $(N-1)$ strips in the phase space.  The well ordered
symbols $w_t \in \{ 0,1,2\ldots ,(N-2) \}$ are constructed from the
anticlockwise enumeration of the disks $s_t \in \{ 1,2,\ldots ,N\} $
using the algorithm 
\begin{eqnarray}
  v & = s_t-s_{t-1} &  \label{a_N_disk}\\
\mbox{if} & v_t<1
 & \mbox{then}\;\;\; v_t = v_t + N  \nonumber \\
w_t & = &  \left\{ \begin{array}{l} 
			v_t - 1\;\;\; \mbox{if}\;\;\; t\mbox{\ odd} \\ 
			N - v_t -1 \;\;\; \mbox{if}\;\;\; t\mbox{\ even} \\ 
                   \end{array}  \right. \nonumber
\end{eqnarray}
and the opposite ordered symbols are $w_t' = N - w_t - 2$.  
When $N=3$ this is the same algorithm as (\ref{a_3_disk}), and for
$N=4$ this is the algorithm in ref.\rf{hansen}.
>From $w_t$ we construct base $(N-1)$ symbolic coordinates
$\gamma = 0.w_1w_2w_3\ldots = \sum_{t=1}^{\infty} w_t / (N-1)^t$ and
$\delta = 0.w_0w_{-1}w_{-2}\ldots = \sum_{t=1}^{\infty} w_{1-t} / (N-1)^t$
where $0\leq \gamma,\delta \leq 1$, and in a analog way $\gamma'$
and $\delta'$.

The reduction to the fundamental domain in the symbol plane is the
same as for 3 disk.  If $N$ is even there is one period 2 orbit with 
only one representation in the symbol plane 
$(\gamma,\delta ) = (\gamma',\delta') = (1/2,1/2)$.


\subsection{$N$ disks with a center disk}

Let the billiard be a configuration of $N$ disks on a
large circle as in the billiard above and in addition one disk 
in the center of this large circle.
The radius of the disks are 1 and the distance between neighbor 
disks are $r$. The disks on the circle is enumerated
anticlockwise from 1 to N and a bounce off the disk in the center is
given the symbol $(N+1)$.
Figure~\ref{f_6_1_config} shows this configuration 
with $N=6$.  If the number of disks on the large circle is even, then, 
because of the disk in the center, a point particle can not bounce
between two disks opposite to each other on the large circle.  
>From the $(N+1)$ symbols of the disks, we get
$(N-1)$ well ordered symbols, and $\Lambda^+_1$ consists of $(N-1)$ 
strips in the phase space. 
With $N$ (even) disks on the large circle and disk number
$(N+1)$ in the center of the large circle the algorithm defining the 
well ordered symbols $w_t
\in \{ 0,1,2\ldots ,(N-2) \}$  is
\begin{equation}
 \begin{array}{lll}
\mbox{if} & s_t = (N+1)
 & \mbox{then}\;\;\; w_t = (N-2)/2  \\
\mbox{else if} &s_{t-1}\neq (N+1)  & \mbox{then}    \\
   & v_t  = s_t-s_{t-1} &    \\
   &\mbox{if}\;\;\;   v_t<1
    & \mbox{then}\;\;\; v_t = v_t + N    \\
    & w_t \;\;\; = &  \left\{ \begin{array}{l} 
			v_t - 1\;\;\; \mbox{if}\;\;\; t\mbox{\ odd} \\ 
			N - v_t -1 \;\;\; \mbox{if}\;\;\; t\mbox{\ even} \\ 
                   \end{array}  \right. \\
\mbox{else if} &s_{t-1} = (N+1)  & \mbox{then}    \\
   & v_t = s_t-s_{t-2} &   \\
   &\mbox{if}\;\;\;   v_t<-(N-2)/2
    & \mbox{then}\;\;\; v_t = v_t + N    \\
   &\mbox{if}\;\;\;   v_t>(N-2)/2
    & \mbox{then}\;\;\; v_t = v_t - N    \\
    & w_t \;\;\; = &  \left\{ \begin{array}{l} 
			(N-2)/2 + v_t\;\;\; \mbox{if}\;\;\; t\mbox{\ odd} \\ 
			(N-2)/2 - v_t\;\;\; \mbox{if}\;\;\; t\mbox{\ even} \\ 
                   \end{array}  \right. 
 \label{a_N_1_disk_even}
 \end{array}
\end{equation}

The configuration with $N=6$ can be looked at as a first step toward a 
description of the Lorentz gas\rf{lorentz,sina70}, 
a triangular lattice with 
a hard disk in each lattice point and a point particle scattering in
the lattice.

If the number of disks $N$ on the large circle is odd, a point particle 
can reach
all other disks after bouncing off one disk when the disks are sufficiently 
separated.  The algorithm giving the
symbols $w_t \in \{ 0,1,2\ldots ,(N-1) \}$ is
\begin{equation}
 \begin{array}{lll}
\mbox{if} & s_t = (N+1)
 & \mbox{then}\;\;\; w_t = (N-1)/2  \\
\mbox{else if} &s_{t-1}\neq (N+1)  & \mbox{then}     \\
   & v_t  = s_t-s_{t-1} &     \\
   &\mbox{if}\;\;\;   v_t<1
    & \mbox{then}\;\;\; v_t = v_t + N     \\
    & w_t \;\;\; = &  \left\{ \begin{array}{l} 
			v_t - 1\;\;\; \mbox{if}\;\;\; t\mbox{\ odd} \\ 
			N - v_t -1 \;\;\; \mbox{if}\;\;\; t\mbox{\ even} \\ 
                   \end{array}  \right. \\
   &\mbox{if}\;\;\;   w_t>(N-1)/2
    & \mbox{then}\;\;\; w_t = w_t + 1     \\
\mbox{else if} &s_{t-1} = (N+1)  & \mbox{then}     \\
   & v_t = s_t-s_{t-2} &    \\
   &\mbox{if}\;\;\;   v_t<-(N-1)/2
    & \mbox{then}\;\;\; v_t = v_t + N     \\
   &\mbox{if}\;\;\;   v_t>(N-1)/2
    & \mbox{then}\;\;\; v_t = v_t - N     \\
    & w_t \;\;\; = &  \left\{ \begin{array}{l} 
			(N-1)/2 + v_t\;\;\; \mbox{if}\;\;\; t\mbox{\ odd} \\ 
			(N-1)/2 - v_t\;\;\; \mbox{if}\;\;\; t\mbox{\ even} \\ 
                   \end{array}  \right. 
 \label{a_N_1_disk_odd}
 \end{array}
\end{equation}

\subsection{Orientation exchange at a bounce}

One can explain why all the algorithms reverse the ordering of symbols
at each bounce by a detailed description of a bounce.  Assume two
particles move along two arbitrarily close parallel lines and bounce off the
dispersing (concave) wall at an angle $\phi \neq 0$ as in
fig.~\ref{f_parallel_concave}.  The particle that first
hits the
wall bounces off the wall along a new direction.  The second particle crosses
the first outgoing line and then bounces off the wall.  The two outgoing
lines cannot cross each other as long as the wall is concave or
straight.  The particle which is on the left side before the bounce, moves on
the right side after the bounce, and 
this causes a change in the orientation at the
bounce.  If the angle $\phi = 0$, the orbits do not cross but because
the velocity is reversed, the orbit that was on the left side before
the bounce is 
on the right side after the bounce.

For a bounce off a focusing (convex) wall as in
fig.~\ref{f_parallel_convex}, the 
two parallel orbits cross twice if $\phi 
\neq 0$.  There is one crossing after the first particle has bounced
off the wall and one crossing after the second particle has bounced
off the wall.  The particle which is on the 
left side before the bounce is on the left side also after the bounce,
and the orientation 
does not change.  This is true also for $\phi = 0$ when the orbits
cross once after the bounces.  These bounces take place e.g. for
a point particle moving in the stadium billiard.  Orbits bouncing only off 
the semicircles parts of the stadium walls can therefore be described 
by an ordered 
rotation number\rf{meiss}.  The description of all orbits in the stadium 
billiard in well ordered symbols are given in ref.\rf{hansen_b}.


\subsection{A condition for sufficiently separated disks}

\newtheorem{condi}{Condition}

We want a condition on the geometrical construction of the
non-wandering set to distinguish between sufficiently separated disks
and the case of pruning.  We find that the disks are sufficiently
separated if condition 1 below is true.

The definition of $\Lambda^+_1$ in the general case is as follows:
$\Lambda^+_1$ consists of $M$ strips and each strip $m^+$ is the union
of points
$(x,\phi )$ from which a straight line starting at $x$ with angle $\phi$
hits a particular disk.  As a straight line may go through other disks
the $M$ strips are not necessarily disjoint.  In the same way we define
$\Lambda^-_1$ as the $M$ strips where each strip $m^-$ is the union of 
points $(x,\phi)$ where a line from point $x$ with angle $-\phi$ hits a
particular disk.  We call the intersection of a strip $m^+$ and a  strip
$m^-$ a rectangle.  The set $\Lambda_1 = \Lambda^+_1 \cap \Lambda^-_1$
then consists of $M^2$ rectangles not necessarily disjoint.  The
construction of $\Lambda_T$ with $T>1$ follows from demanding that the
outgoing angle at a bounce is equal to the incoming angle, but allowing
the straight lines to go through a disc.  Then $\Lambda_T$ consists of
$M^{2T}$ rectangles.
\begin{condi}
   There exists a number $0<T<\infty$ such that $\Lambda_T$
   consists of $M^2$ disjoint areas where each area is inside one of
   the $M^2$ rectangles of $\Lambda_1$.
\end{condi}

The iteration of the $M^2$ disjoint areas corresponds to one more bounce and
gives that each of the $M^2$ disjoint areas contains $M^2$ new disjoint
areas.  The $M^4$ rectangles of $\Lambda_2$ then contains $M^4$ disjoint
areas.  By induction we find that $\Lambda_{T+T'}$ gives $M^{2T'}$
disjoint areas inside the $M^{2T'}$ rectangles of $\Lambda_{T'}$.  From this
it follows that even if the rectangles of $\Lambda_{T'}$ overlap each
other, the part of the non-wandering set belonging to the rectangles  do
not overlap.  A symbol string $(w_{-T'+1}w_{-T'+2}\ldots w_{T'-1}w_T')$,
with $w_t\in \{ 0,1,\ldots , (M-1)\} $ uniquely correspond to one
rectangle when describing the part of the non-wandering set in this
rectangle.  It then follows that the disks are sufficiently separated.

If there does not exist a $T$ according to the condition it may be that
an infinitesimal change of the parameter gives a $0<T<\infty$ and this is the
critical parameter value where pruning starts.  If non of the above is
true, then the rectangles always overlap and there is pruning in the
system and not all symbol strings corresponds to an orbit.


\section{Pruning}
                                                             
When the distance $r$ between the disks is small, a subset of the orbits are
forbidden (not admissible).  We conjecture that
only two kinds of forbidden orbits exist in the dispersing billiards; 
orbits passing through a
forbidden region (e.g. through a disk) and orbits going into a forbidden region
and bouncing off the wall from the focusing side.  Figure
~\ref{f_pruning_orbits} a) shows a part of two {\em legal} orbits; one
passing the dispersing wall and one bouncing off the wall on the
dispersing side.  Figure~\ref{f_pruning_orbits}  c)
shows a part of two {\em forbidden}s orbits; one passing through the
dispersing wall and the other going into the forbidden region and bouncing 
off the wall from the focusing side.
Figure~\ref{f_pruning_orbits} b) shows that the limit orbit
of both orbits is a line tangential to  the disk.  An orbit is therefore 
pruned together with at least one other orbit.

The orbits bouncing off the wall from the focusing side are orbits
with  $|\phi | > \pi /2$ in the phase space. 
In the phase space for sufficiently separated disks all parts of
$\Lambda$ have angle $|\phi |<\pi /2$.  When $r$ decreases,
some points in $\Lambda$ move closer to the lines $|\phi | =
\pi/2$ and the pruning
starts at $r_c$ when the points in $\Lambda$ with largest value of 
$|\phi |$ reach these lines.

When $r\rightarrow r_c$ with $r > r_c$, the value of $T$ in condition 1
tends to
$\infty$.  For $r < r_c$ the different rectangles of $\Lambda_t$ always
have some overlapping.  This overlapping corresponds to forbidden 
regions in the
symbol plane.  The forbidden regions that are the simplest to describe
are the two regions consisting of points in $\Lambda$ with $|\phi | > \pi/2$. 
The borders of these
regions in phase space are the lines $|\phi | = \pi /2$.  In symbol
plane the two regions correspond to two areas, one in the upper left 
corner and the other in the bottom right corner.
The border is the symbolic representation of all points in $\Lambda$
with $\phi = \pi/2$.   To obtain this border, we scan through
$x$ values with the angle $\phi = \pm\pi/2$ and if $r>2$
(an open system) in numerical work we only keep points
bouncing more than 20 bounces both in the future and in the past.

As both $\gamma $ and $\delta$ increase with the
value of $x$,
the border is monotonously increasing in the symbol plane.

The forbidden regions described above contain only the forbidden
orbits bouncing from the focusing side and not the orbits passing through  
a forbidden region.  We choose
to take the pre--image of this described region as our primary
forbidden region and the pre-image of the border as our pruning
front.  The other family of forbidden orbits, that is orbits passing through
a disk, is an other primary forbidden region and has a second pruning front
that is very similar to the first pruning front.  An orbit on this
second pruning front has the same symbol sequence as an orbit on the
first pruning front but without the one symbol for the tangential
bounce.  The two primary forbidden regions have a straight line as a
common border and this line corresponds to a gap between two bands in
$\Lambda^+_1$ for sufficiently separated disks.  We consider the two 
primary forbidden regions to be bounded by one pruning front.  
This region can be
described as the primary overlapping in $\Lambda$. 
In one of the examples we will show that there may be more than one of
these primary forbidden regions, but all are constructed in the same
way.

The iteration of the points in the symbol plane is a shift operation of
the symbols.  This shift operation is slightly more complicated for well
ordered symbols than for the original disk symbols, see ref.\rf{troll}.
The primary forbidden region can by the shift operation be mapped
forward and backward in time.  The union of all images and preimages 
of the primary
forbidden region is dense and takes the full measure in the symbol plane; 
Their complement, the union of the legal orbits has measure zero and is 
a Cantor set in the symbol plane.

>From the primary forbidden region we can read off all the finite
sequences that can make an orbit forbidden.  If the well ordered
alphabet has $M$ symbols, we divide the symbol plane into rectangles
with side length $M^{-k}$.  Each rectangle has a unique labeling 
$w_{-k+1}w_{-k+2}\ldots w_{k-1}w_k$ and corresponds to one disjoint area in 
the phase space of 
a sufficiently separated disk system.  If the  rectangle is inside the
forbidden region, the symbol string  $w_{-k+1}w_{-k+2}\ldots w_{k-1}w_k$
is forbidden in the alphabet.  By excluding all the forbidden
symbol strings for a given number $k$ we can redefine the symbolic
dynamics in terms of a new alphabet with a new grammar, or in terms
of a finite Markov graph\rf{cvitanovic,grassberger}.  
This is an approximation of
order $k$ to the correct complete alphabet describing the billiard,
and this gives an approximation to a Markov partition of the
billiard.   
The implementation of such approximations in practice will be treated 
elsewhere.


\subsection{Pruning for 3 disk billiard}

The pruning in the 3 disk billiard starts when the points in $\Lambda $ with 
the largest value of $|\phi |$ reach the value $|\phi | = \pi /2$.
These outermost points in $\Lambda$ are the hetroclinic points created by the 
crossing of the unstable (stable) manifold of the fixed point $W=\overline{0}$
and the stable (unstable) manifold of the fixed point $W=\overline{1}$.
The numerical value of the critical distance is $r_c = 2.04821419\ldots$.
The pruning front for 3 disks with $r<r_c$ is numerically determined 
by choosing
points on disk 2  and starting the orbit by glancing off the disk at point
$x$ with the angle $\phi = \pm\pi/2$.
The first bounce in  one direction is off disk 1 and in the other
direction the bounce is off disk 3. Let $s_0 = 1$, and a point on the first
pruning front is determined by choosing $s_1=2$ and $s_2=3$. 
The other values of $s_t$ are obtained from the numerical bouncing with 
starting point  $x$ on disk 2.
To find a point on the second pruning front
we do not include the bounce in disk 2 but assume the orbit grazes this 
disk.  This gives the symbols $\tilde{s}_t = s_t$ for $t\leq 0$ and 
$\tilde{s}_t = s_{t+1}$ for $t>0$.
The points $(\gamma , \delta )$ are computed from the symbol sequences of 
orbits with starting point $x$ between $2 \pi /3$ and $\pi$.  Plotting 
the points $(\gamma,\delta)$ in the symbol plane gives
fig.~\ref{f_3_pruning_front} for parameter $r=2$ (touching disks). The
pruning front is a monotonously increasing line from $(0.375,0)$ to
$(0.5,0.5)$ and a monotonously decreasing line from  $(0.5,0.5)$ to
$(0.75,0)$.

A numerical test of this pruning front can be performed by starting one
orbit at a random point in the billiard, letting it bounce in the billiard
for some time and calculate the point $(\gamma,\delta)$ for each bounce.  
In fig.~\ref{f_3_long_orbit}~a) each bounce for an
orbit with $10^6$ bounces is marked by a point in the symbol
plane.  As expected, the primary forbidden region is white in
fig.~\ref{f_3_long_orbit}~a) as there are no bounces with these
$(\gamma,\delta)$ values.  In addition, there are 3 copies 
of this white
region; the points $(\delta,\gamma)$ that are a result of the time
reversion symmetry ($t \rightarrow -t$) and the points
$(1-\gamma,1-\delta)$ and $(1-\delta,1-\gamma)$ that are a consequence
of the symmetry between odd and even in algorithm (\ref{a_3_disk}).
This last symmetry is the results of that our symbol plane includes two
fundamental domains.
All other white areas in fig.~\ref{f_3_long_orbit}~a) are images or 
preimages of one primary forbidden region, where the iteration is a shift
operation on the symbol string.

Fig.~\ref{f_3_pruning_front} and fig.~\ref{f_3_long_orbit}~b)  are
magnifications of the symbol plane showing respectively the pruning 
front and the distribution of
points of a long orbit.  The long chaotic orbit visits points close
to both pruning fronts but the probability is different as visualized
in fig.~\ref{f_3_long_orbit}~b).  The orbit is often close to grazing a
disk but very seldom does the orbit bounce with angle close to
$\pi/2$.  The tangential bounce is very unstable and this gives low
probability density in the symbol plane to these bounces.  The stability
is however not 
sensitive to how close an orbit is to graze a wall, and the 
probability density in the symbol plane approaches the second pruning 
front rather uniformly.

One can convert the topological pruning front in
fig.~\ref{f_3_pruning_front} into a symbolic description in different
ways. The following method overcounts the number of allowed orbits.

In fig.~\ref{f_3_pruning_front} rectangles with side length $2^{-7}$ are
plotted together with the pruning front.  
A rectangle with side length $2^{-k}$ in the symbol plane corresponds to a
symbol string of length $2k$ in the symbolic description.  Each
rectangle in fig.~\ref{f_3_pruning_front} is identified with a symbol
string  $w_{-6}w_{-5}\ldots w_6w_7$.
If a rectangle is completely inside the primary
forbidden region then the corresponding symbol string is forbidden.  
The length 14 symbol
strings $w_{-6}w_{-5}\ldots w_6w_7$ that are forbidden for 3 touching disks
are listed in
column 2 in table~\ref{t_3_forbidden_symb}.
Finding the completely forbidden strings of length $2k$ for all 
$k\in \{ 1,2,\ldots \} $ gives a complete list of forbidden orbits.
Table~\ref{t_3_forbidden_symb} gives this list for $k\leq 9$ when $r=2$.
The shortest forbidden symbol string is of length 12 ($k=6$).
In addition to the symbol strings $w$ in table~\ref{t_3_forbidden_symb} 
the $w'$ strings (obtained by letting $0\rightarrow 1$ and $1\rightarrow
0$) and the 
time reversed strings of $w$ and $w'$ have to be included as forbidden orbits.

It is possible to include in the table of forbidden strings 
the finite symbol strings
describing rectangles that are just partly inside the primary forbidden
region.   This gives an undercounting of the legal orbits.  It is
also possible to apply a combination of these two methods.  Other
strategies are using some periodic orbits close to the pruning front as
an approximation, or to use some number of points that are exactly on
the pruning front.  It is presently not clear which approach yields
the most convergent sequence of approximate grammars.

The pruning front is not a continuous line in the symbol plane.  The
images and preimages of the pruning front cross the primary forbidden
region and create open intervals in the pruning front.  The result is
that the pruning front itself is a Cantor set.  
The open intervals do not lead to any
problems in converting the pruning front into a table of forbidden
orbits.  Any monotone curve in the interval, e.g. a straight line,
gives a correct result.

The pruning front for an open 3 disk system with $r=2.02$ is plotted in figure
\ref{f_3_pruning_front_2} and for a billiard with overlapping disks with 
$r=1.97$ in figure
\ref{f_3_pruning_front_3}.  As expected the figures show that the 
primary pruned region increases when the distance $r$ decreases.
The structure of the pruning front does not
change dramatically when the distance changes.  
In the billiard with overlapping disks, $r<2$, the point particle can not
bounce infinitely many times in a
corner of the billiard and this gives an upper and lower bound on $\gamma$ and 
$\delta$: $0<\beta \leq \gamma , \delta \leq (1-\beta ) < 1$.
These strips of width $\beta $ and their images and preimages are also 
forbidden regions.
Figure~\ref{f_3_long_r_19} shows a long orbit with $r=1.9$. A large 
area in the lower left corner and in the upper right corner are pruned by these 
strips.


\subsection{Pruning for 4 disk billiard}

The critical parameter value where the pruning start for the 4 disk system
is $r_c = 2.20469453\ldots$.
The pruning front is similar to the pruning front for the 3 disk billiard.
This billiard and the hyperbola billiard are discussed in detail in 
ref.\rf{hansen}.

\subsection{Pruning for a $6+1$ disk system}

The billiard with 6 disks on a large circle and one disk in the center 
has pruned
orbits when the distance between the disks is less than 
$r_c=3.59148407\ldots$. 
Algorithm~(\ref{a_N_1_disk_even})  with $N=6$ gives the well ordered
symbols for these system.  Figure~\ref{f_6_1_config} shows the 7 disks and 
a part of one orbit that is grazing disk 7 (the center disk) and the
symbolic coordinate $(\gamma,\delta)$ of this orbit is on
the pruning front created by this center disk.  The symbol plane used
corresponds to the phase space of disk 1.   In this system the disk~7
has a different phase space and symbol plane and the pruning front obtained
is only valid for one of the disks on the large circle.
There are two pruning fronts in the symbol plane; an orbit bouncing off
disk 1 can graze either disk 7 as in
figure~\ref{f_6_1_config} or it can graze disk 2.  The primary pruned
region in the symbol plane is limited by these pruning fronts as shown in
figure~\ref{f_6_1_prun_front} for parameter $r=2.2$. One orbit 
grazes both disk~2 and disk~7 and this orbit gives a point
connecting the two pruning fronts.

The primary pruned region is converted to a list of forbidden symbol strings  
similarly to the 3 disk billiard.
Rectangles of size $5^{-k}$ correspond to symbol strings 
$w_{-k+1}w_{-k+2}\ldots w_{k-1}w_k$ and the rectangles in the primary pruned 
region correspond to forbidden strings. 
Table~\ref{t_6_1_forbidden} lists all the forbidden orbits of 
length 2 and 4.

 
\section{Conclusions}

We have shown that it is possible to define a well ordered covering
alphabet for different concave billiard systems with a finite number
of concave walls.  For these systems the primary forbidden region is 
a simple connected region in
a symbol plane.  The border of this region is the
pruning front, and we have found this pruning front numerically for 
the 3 disk system with different parameter values.  The 
pruning front is also obtained for a more complicated billiard with 7 disks.
The general method to construct a symbolic alphabet ordered the
same way as the ordering of orbits in the phase space is discussed.  
A well ordered alphabet is necessary 
when finding the pruning front because, for any other alphabet the primary 
pruned region is a very complicated region.

We show how the topological information, the pruned region, can be 
converted into symbolic dynamics, a list of
forbidden symbol strings.  The tables list some of such
forbidden strings.

The method can easily be applied to other finite concave billiards with
similar symmetries as those 
considered in this article.  Infinite billiards like the Bunimovich stadium 
billiard, the Sinai billiard and the Lorentz gas systems are more complicated 
because it is difficult to obtain a well ordered covering alphabet which 
is not pruned.  These systems will be treated elsewhere. 
A smooth potential described by the same symbolic dynamics as a disk
billiard system\rf{DR,wintgen} may have a pruning front and will also be
treated in a similar way.  It is however problems in defining a bounce 
in the smooth potential and this is presently under investigation.

\newpage

\begin{enumerate}

\fg{f_3_config}
The three disks in configuration space enumerated anticlockwise.

\fg{f_6_1_config}
The 6 + 1 disks, enumerated anticlockwise.  $r=2.2$.
The orbit is tangent to disk 7, and therefore belongs to the pruning 
front for the system.

\fg{f_parallel_concave}
Show how two parallel orbits change orientation when bouncing in a 
concave wall.

\fg{f_parallel_convex}
Show that two parallel orbits do not change orientation bouncing in
a convex wall.

\fg{f_pruning_orbits}
The allowed, boundary and forbidden orbits:
\ a) Two legal orbits. \ b) Two orbits on the pruning front.
\ c) Two forbidden orbits.

\fg{f_3_long_orbit}
The symbol plane picture of an orbit bouncing $5\cdot 10^6$ times in the
closed 3 disk system with touching disks (center distance $r=2$).  
Each bounce corresponds to a point
in the plane.  \ a) the whole symbol plane.  \ b) magnification of
a part of the symbol plane showing the primary pruned region.

\fg{f_3_pruning_front}
A magnification of the pruning front for 3 disks with $r=2$,
with line mesh spacing at $ 2^{-7}$.  
The symbols give the symbolic past and
the symbolic future of the rectangles.

\fg{f_3_pruning_front_2}
The pruning front for 3 disks with $r=2.02$.

\fg{f_3_pruning_front_3}
The pruning front for 3 disks with $r=1.97$

\fg{f_3_long_r_19}
An  orbit bouncing $2\cdot 10^6$ times in the
closed 3 disk system with center distance $r=1.9$.

\fg{f_6_1_prun_front}
The pruning front for 6+1 disks,  $r=2.2$.

\end{enumerate}

\newpage


{\bf Acknowledgements: }
The autor is grateful to 
P.~Cvitanovi\'c, T. Schreiber, F. Christiansen, P. Dahlqvist, M. Sieber
and Z. Kaufmann for 
discussions.
The autor thanks the Norwegian Research Council for support.


\renewcommand{\baselinestretch} {1}

\newpage



\begin{table}[p]
\[
\begin{array}{|l|l|}
\hline
s_{0}s_{1}s_{2}s_{3}s_{4} &  w_{1}w_{2}w_{3}w_{4} \\
\hline
12121 & 0000 \\
12123 & 0001 \\
12132 & 0010 \\
12131 & 0011 \\
12313 & 0100 \\
12312 & 0101 \\
12321 & 0110 \\
12323 & 0111 \\
13232 & 1000 \\
13231 & 1001 \\
13213 & 1010 \\
13212 & 1011 \\
13121 & 1100 \\
13123 & 1101 \\
13132 & 1110 \\
13131 & 1111 \\
\hline
\end{array}
\]

\caption{
\label{t_3_symbols}
The table shows the ordering of 4 bounces of a particle
starting on disc 1 at time 0.  The symbol $s_t$ is 
the number of the disc, while $w_t$ is new symbols 
reflecting the ordering in phase space.
}

\end{table}


\begin{table}[p]

\[
\begin{array}{|l|l|l|}
\hline
\_ w_{-5}\ldots w_6\_ & \_ w_{-6}\ldots w_7\_ & \_ w_{-7}\ldots w_8\_ \\
\hline
000000\cdot 100000
 & 0000000\cdot 1000011 & 00000000\cdot 10001010 \\
 & 0000000\cdot 1000010 & 00000000\cdot 10001001 \\
 & 0100000\cdot 1000001 & 00000000\cdot 10001000 \\
 & 0100000\cdot 1000000 & 01000000\cdot 10000101 \\
 & 1100000\cdot 1000000 & 01000000\cdot 10000100 \\
 & 0010000\cdot 1000000 & 11000000\cdot 10000100 \\
 & 1000000\cdot 0111111 & 01100000\cdot 10000010 \\
 & 0000000\cdot 0111111 & 11100000\cdot 10000010 \\
 & 0000000\cdot 0111110 & 01010000\cdot 10000001 \\
 & 0100000\cdot 0111111 & 01010000\cdot 10000000 \\
 &                      & 11010000\cdot 10000000 \\
 &                      & 00110000\cdot 10000000 \\
 &                      & 10110000\cdot 10000000 \\
 &                      & 01110000\cdot 10000000 \\
 &                      & 11110000\cdot 10000000 \\
 &                      & 00001000\cdot 10000000 \\
 &                      & 01010000\cdot 01111111 \\
 &                      & 10010000\cdot 01111111 \\
 &                      & 00010000\cdot 01111111 \\
 &                      & 11100000\cdot 01111111 \\
 &                      & 01100000\cdot 01111111 \\
 &                      & 01000000\cdot 01111101 \\
 &                      & 00000000\cdot 01111011 \\
\hline
\end{array}
\]

\caption{
\label{t_3_forbidden_symb}
The forbidden orbits in 3 discs, $r=2$. ``Under counting'' approximation to
level 9.
}

\end{table}


\begin{table}[p]

\[
\begin{array}{|l|l|l|}
\hline
\_ w_{-8}\ldots w_9\_ & & \\
\hline
000000000\cdot 011110011
 &  110101000\cdot 100000000  
   &  001000000\cdot 100001110  \\
000000000\cdot 011110100
 &  010101000\cdot 100000000  
   &  001000000\cdot 100001111  \\
000000000\cdot 011110101
 &  100101000\cdot 100000000  
   &  110000000\cdot 100010000  \\
100000000\cdot 011110101
 &  000101000\cdot 100000000  
   &  010000000\cdot 100010000  \\
010000000\cdot 011110111
 &  111001000\cdot 100000000  
   &  010000000\cdot 100010001  \\
001000000\cdot 011111000
 &  011001000\cdot 100000000  
   &  000000000\cdot 100010110  \\
001000000\cdot 011111001
 &  101001000\cdot 100000000  
   &  000000000\cdot 100010111  \\
011000000\cdot 011111011
 &  001001000\cdot 100000000  
   &  000000000\cdot 100011000  \\
111000000\cdot 011111011
 &  110001000\cdot 100000000  
   &  000000000\cdot 100011001  \\
000100000\cdot 011111011
 &  010001000\cdot 100000000 & \\
001100000\cdot 011111101
 &  000110000\cdot 100000010 & \\
101100000\cdot 011111101
 &  111010000\cdot 100000010 & \\
011100000\cdot 011111101
 &  011010000\cdot 100000010 & \\
111100000\cdot 011111101
 &  011010000\cdot 100000011 & \\
011010000\cdot 011111110
 &  010010000\cdot 100000100 & \\
011010000\cdot 011111111
 &  100010000\cdot 100000100 & \\
111010000\cdot 011111111
 &  000010000\cdot 100000100 & \\
000110000\cdot 011111111
 &  000010000\cdot 100000101 & \\
100110000\cdot 011111111
 &  001100000\cdot 100000110 & \\
010110000\cdot 011111111
 &  010100000\cdot 100001000 & \\
110110000\cdot 011111111
 &  100100000\cdot 100001000 & \\
001110000\cdot 011111111
 &  000100000\cdot 100001000 & \\
101110000\cdot 011111111
 &  000100000\cdot 100001001 & \\
011110000\cdot 011111111
 &  111000000\cdot 100001010 & \\
111110000\cdot 011111111
 &  011000000\cdot 100001010 & \\
000001000\cdot 011111111
 &  001000000\cdot 100001100 & \\
100001000\cdot 011111111
 &  001000000\cdot 100001101 & \\
\hline
\end{array}
\]

Table 2: Continue.

\end{table}



\begin{table}[p]

\[
\begin{array}{|l|l|l|l|l|l|l|}
\hline
\_ w_{0}w_{1} \_ & \_ w_{-1}&\ldots w_2 \_ &&&& \\
\hline
4\cdot 1 & 04\cdot 20 & 43\cdot 20 & 32\cdot 11 & 12\cdot 14 & 11\cdot 14 
		& 00\cdot 14 \\ 
3\cdot 1 & 14\cdot 20 & 42\cdot 20 & 32\cdot 12 & 02\cdot 13 & 21\cdot 14 
		& 10\cdot 14 \\
         & 24\cdot 20 & 32\cdot 20 & 32\cdot 13 & 02\cdot 14 & 31\cdot 14 
		& 20\cdot 14 \\
         & 34\cdot 20 & 42\cdot 10 & 32\cdot 14 & 01\cdot 13 & 41\cdot 14 
		& 30\cdot 14 \\
         & 44\cdot 20 & 42\cdot 11 & 22\cdot 11 & 11\cdot 13 & 00\cdot 13 
		& 40\cdot 14 \\
         & 03\cdot 20 & 42\cdot 12 & 22\cdot 12 & 21\cdot 13 & 10\cdot 13 
		& 10\cdot 10 \\
         & 13\cdot 20 & 42\cdot 13 & 22\cdot 13 & 31\cdot 13 & 20\cdot 13 
		& 00\cdot 10 \\
         & 23\cdot 20 & 42\cdot 14 & 22\cdot 14 & 41\cdot 13 & 30\cdot 13 
		& 00\cdot 11 \\
         & 33\cdot 20 & 32\cdot 10 & 12\cdot 13 & 01\cdot 14 & 40\cdot 13 
		&  \\
\hline
\end{array}
\]

\caption{
\label{t_6_1_forbidden}
The forbidden orbits in 6+1 discs, $r=2.2$. ``Under counting'' 
approximation to level 2.
}

\end{table}

\end{document}